\documentclass{amsart}
\usepackage{amssymb}

\usepackage{amsfonts}
\usepackage[english]{babel}

\usepackage[T1]{fontenc}
\usepackage{latexsym}

\usepackage{color}

\newtheorem{theorem}{Theorem}[section]

\newtheorem{proposition}[theorem]{Proposition}
\newtheorem{corollary}[theorem]{Corollary}
\theoremstyle{definition}
\newtheorem{remark}[theorem]{Remark}

\newcommand{\zfrac}[2]{#1/#2}

\usepackage{fancyhdr}
 \usepackage{lastpage}
\author{Jacek
Jakubowski and
 Maciej Wi\'sniewolski }\thanks{Research partially supported by Polish MNiSW grant N N201 547838.}

\title[
Linear SV
model]{
Linear stochastic volatility models}

\begin{document}

\maketitle
\begin{center}
{\small
 Institute of Mathematics, University
of Warsaw \\
  Banacha 2, 02-097 Warszawa, Poland \\
e-mail: {\tt jakub@mimuw.edu.pl } \\
 e-mail:  {\tt M.Wisniewolski@mimuw.edu.pl} }

\end{center}

\begin{center}
First version  September 25, 2009 \ \   This version \today
\end{center}
\begin{abstract}
In this paper we investigate general  linear stochastic volatility models with correlated Brownian
noises.  In such models the asset price satisfies a linear SDE with
coefficient of linearity being the volatility process. This class contains among others Black-Scholes model,  a log-normal stochastic
volatility model and Heston stochastic volatility model.
For a linear stochastic volatility model we derive  representations for the probability density
function of the arbitrage price of
a financial asset and the prices of European call and put options.
 A closed-form formulae for the density function and the prices of European call and put options
are given for log-normal stochastic volatility model. We also obtain present some  new results for
Heston and extended Heston stochastic volatility models.

\end{abstract}

\noindent
\begin{quote}


\end{quote}

\section{Introduction}
The famous Black-Scholes model with its relatively stringent
assumptions does not capture many phenomena of modern financial
markets. A prime example  is the
stochastic nature of the financial asset's volatility, called volatility
smile (see for example Hull and White \cite{Hul}). In recent
years many stochastic volatility models have been introduced and
developed. However, making the volatility
stochastic   complicate the models considerably
(see for example Rebonato \cite{RE}).
It is not our aim
 to review the broad range of stochastic volatility models.
 We focus on and develop the idea of modeling stochastic
volatility in the simplest possible but effective way.  SABR is an
excellent example of a model complex in  nature but simple
in form. This well known and celebrated model, introduced in
2002 by Hagan et al. \cite{Hag1}, has
been
effectively used and investigated by market practitioners.
It turned out, soon
after its introduction,  that it is more effective than
Black-Scholes and local volatility models.
The key idea in SABR is
to make stochastic volatility a simple stochastic process and then
shift the difficulty of finding the financial asset's distribution to
the level of finding the  distribution of the diffusion describing the asset
price.  Determining  closed formulae for the asset price
distribution in a SABR model remains, in general, an unsolved problem (as far as the
authors know).
The task of determining
closed formulae for the  probability distribution in a SABR model with the parameter
beta equal to one, called a log-normal stochastic volatility model, has
 been investigated by Maghsoodi \cite{Mag}, \cite{Mag1}. In
this case
it is possible to write out the solution of the model, i.e.  the stochastic process
representing the  asset price, as the exponential  of a linear
combination of  functionals of a pair of correlated Brownian motion.
Maghsoodi used the techniques of changing time and changing  measure to find
the joint density function of these functionals. The same techniques
had been used earlier by Yor in the problem of
valuation of Asian options (see
\cite{YOR}).
 However,  Maghsoodi
 did not  mentioned that the asset price loses the
martingale   property in a log-normal stochastic
volatility model in the case of positive correlation between the
asset price and its volatility.

In our work we  reverse the idea of the SABR
model and continue the line of research of Hull and White \cite{Hul} followed
also by Romano and Touzi \cite{RT}  as well as by Leblanc
\cite{Leb}. We shift the complicated nature of the model to the
level of the process representing volatility, keeping the
diffusion of the asset price relatively simple.
 So, we assume that the asset price process $X$ satisfies
$
 dX_t = Y_tX_tdW_t
$
 with $Y$ given by
 $   dY_t = \mu(t,Y_t)dt + \sigma (t,Y_t)dZ_t,
 $
where the processes $W$ and $Z$ are correlated Brownian motions. We call this
model a linear stochastic volatility model.
 We prove that the distribution of the asset price in an arbitrary  linear stochastic volatility model has a density
function and we derive the  representation of that
function (Theorem \ref{tw:densrep}). This representation
depends on some functionals of the process representing volatility, so
 the problem of determining the  asset price distribution
reduces to finding the distribution of a 2-dimensional  functional of
the volatility.
  In Section 3, we point out two nontrivial examples of such models in
which we can benefit from  representations of the asset price
density function.
 The first example
is a log-normal stochastic volatility model which is a SABR
model with beta equal to one
(it is also called the Hull-White model).
We find  closed
formulae for the density function in a log-normal stochastic volatility model using the result of Matsumuoto and Yor
\cite{Mat} who derived the density function for the vector of Brownian
motion with drift and its exponential functional.
  In Section 4 we derive representations for European call and put option
prices in the linear stochastic volatility models.
The  representation for vanilla option prices is
independent of the distribution of the asset price itself.
In particular, this allows us to obtain formulae for the arbitrage prices of
vanilla  options in a log-normal stochastic volatility
model.
    Similar  representations for European call and put
option arbitrage prices in a linear stochastic volatility model have
also been given by Romano and Touzi \cite{RT}, but in a slightly
different context. They considered a slightly different model and
established a set of assumptions under which they obtained
 representation results while proving the convexity of
European call and put options in their setting (also linear in our
sense).
 In particular,  they assumed that the coefficients $\mu$ and $\sigma$
in the definition of $Y$ are bounded. In our work we relax  this
assumption (see Theorem \ref{tw:probrep}). In our examples  the
drift coefficient is not bounded, but the  representation for option prices holds. It should be mentioned that  Leblanc
\cite{Leb} gives the arbitrage price
of call option in a linear stochastic volatility model, with some
concrete examples of volatility, in terms of Laplace and Fourier
transforms.
 Closed formulae for the density function and vanilla option prices in
a stochastic log-normal volatility model are interesting and
important for applications since such models are popular, especially
among the forex exchange options traders (see \cite{Hag1}). Similar
results for log-normal stochastic volatility models were
also presented in \cite{Mag} and \cite{Mag1}.
    In Section 5 we present connections between a distribution of the asset price process and  prices  of put options.
In a linear stochastic volatility model
we represent the distribution of the process $X$ giving the price of the asset
in terms of  prices of put options (see Thm. \ref{DSTR}).
In Corollary \ref{trfun} we find that a Laplace transform of $X_t$ for  $\lambda > 0$ is equal to price of put option with random strike multiplies by constant.
 Next we consider the log-normal stochastic volatility model. We present a relatively simple proof of the fact that
the price process $X$ is a martingale if and only if $\rho \leq 0$.
 As an example we indicate a possible applications of our results to the Hull-White  model.
  Taking the  parameter  $\rho$ calibrated to market prices of the options, we can obtain  the calibrated distribution of the asset price process.
 In Section 6 we consider the Heston and extended Heston volatility models. The first and the most important result, which we present for these models, is that the asset price is always a true martingale under a martingale measure. It is the new result and the
 significant extension of results obtained by Wong and Heide \cite{HW}. These authors assumed, after Heston, the special form of density  of martingale measure and under assumptions concerning the parameters of the model showed that the asset price process is a martingale. In this paper we assume neither some special form of martingale measure nor some additional assumptions about model parameters. We also find the Laplace transform of volatility functional in the  extended Heston model and propose some new approximation method of finding the Laplace transform of vanilla option price.

\section{ Representation of the density function of the asset price
in a linear stochastic volatility model}
\subsection{Linear stochastic volatility models.}

We consider a market defined on a complete probability space
$(\Omega,\mathcal{F},\mathbb{P})$ with filtration $\mathbb{F}=
(\mathcal{F}_t)_{t \in [0,T]}$, $T<\infty$, satisfying the usual
conditions and  $\mathcal{F}=\mathcal{F}_T$. Without loss of
generality we assume the savings account to be constant and
identically equal to one. Moreover, we assume that  the price $X_t$ at time $t$
of the underlying asset  has a stochastic volatility
$Y_t$, and the dynamics of the vector $(X, Y)$ is given by
\begin{align}
    dX_t &= Y_tX_tdW_t, \label{defX} \\
    dY_t &= \mu(t,Y_t)dt + \sigma (t,Y_t)dZ_t, \label{defY}
\end{align}
where $X_0$, $Y_0$ are positive constants, the processes $W,Z$ are
correlated Brownian motions, $d{\left\langle W,Z\right\rangle}_t =
\rho dt$ with $\rho\in(-1,1)$,  and $\mu: \mathbb{R}_+ \times
\mathbb{R}_+ \rightarrow \mathbb{R}$, $\sigma: \mathbb{R}_+ \times
\mathbb{R}_+ \rightarrow \mathbb{R}$ are continuous functions such
that there exists a unique strong solution of \eqref{defY}, which is
positive and $\int_0^T Y_u^2du < \infty$ $\mathbb{P}$-a.s.

Under these
assumptions the process $X$ has the form
\begin{equation}
  X_t = X_0e^{\int_0^tY_udW_u - \zfrac{\int_0^tY_u^2du}{2}} , \label{postacX}
\end{equation}
and this is a unique strong solution of SDE \eqref{defX} on
$[0,T]$. The existence and uniqueness follow
directly from the assumptions on $Y_t$ and the well known properties
of stochastic exponent (see, e.g., Revuz and Yor \cite{RY}). The
process $X$ is a local martingale, so there is no arbitrage on the market so
defined.

We  call this model a {\it linear stochastic volatility model},
because the SDE \eqref{defX} governing the asset price is
linear with respect to the asset price itself with coefficient
being the stochastic volatility $Y$.
Note that the known models such as Black and Scholes model, log-normal stochastic volatility model, Heston model (where $Y^2$ is a CIR process) and Stein and Stein model belong to this class.
\begin{remark} a)
It is worth  mentioning that the constant $\rho$ in the model can be
replaced by a measurable, deterministic function $\rho :
[0,T]\rightarrow (-1,1)$ and  the results of this work remain true with
minor modifications. \\
b) Our standing assumption is $|\rho|<1$. However, our methods allow
 finding the distribution of $X_t$ in  the case $\rho =\pm 1$. Indeed,    we have $W=\pm Z$
in this case and
\begin{equation}
    X_t = X_0e^{\pm\int_0^tY_udZ_u - \zfrac{\int_0^tY_u^2du}{2}} , \label{postacXX}
\end{equation}
so the problem of finding the distribution of $X_t$, for fixed $t$,
reduces  to deriving the distribution of the vector
$(\int_0^tY_udZ_u, \int_0^tY_u^2du)$.
\end{remark}

\subsection{Existence of the density function and its  representation.}

We start with the main theorem of the paper on existence of the
density function of the underlying asset price in a linear stochastic
volatility model, and its  representation.
This
representation allows us to find a closed formula for the  density function
(see examples in the next section), which is  important for
applications (see, e.g.,  Carmona and Durrleman \cite{Car}).
\begin{theorem}\label{tw:densrep}
 Fix $t\in [0,T]$. In a linear stochastic
volatility model  the distribution of  $X_t$ has the
 representation
\begin{equation} \label{eq: nowy}
 \mathbb{P}(X_t \leq r)
 = \mathbb{E}\Phi\bigg(\frac{\ln\frac{r}{X_0}-\mu_Z(t)}{\sigma_Z(t)}\bigg),
\end{equation}
 where $r>0$, $\phi$ is the density function of a standard
Gaussian random variable $N(0,1)$, and
\begin{align} \label{eq:repr_mu}
    \mu_Z(t) &= \rho\int_0^tY_udZ_u - \frac{1}{2}\int_0^tY_u^2du,\\
 \label{eq:repr_si}   \sigma_Z^2(t) &= (1 - \rho^2)\int_0^tY_u^2du.
\end{align}
Moreover, the random variable $X_t$ has
density function $g_{X_t}$, which  has the
 representation
\begin{equation} \label{eq:repr_g}
g_{X_t}(r) =
\mathbb{E}\bigg[\frac{1}{r\sigma_Z(t)}\phi\bigg(\frac{\ln\frac{r}{X_0}-\mu_Z(t)}{\sigma_Z(t)}\bigg)\bigg].
\end{equation}
If
\begin{equation} \label{eq:pod}
\mathbb{E} \Big( \int_0^tY_u^2du \Big)^{-\frac{1}{2}} < \infty ,
\end{equation}
 then  the  density function $g_{X_t}$ is  continuous.
\end{theorem}
\begin{proof}  Notice that we can represent $W$ in the
form
  \begin{equation} \label{eq:repr-W}
  W_t
= \rho Z_t + \sqrt{1-\rho^2}B_t,
\end{equation}
where $(B,Z)$ is the standard two-dimensional Wiener process.
The It\^o lemma applied to \eqref{defX} together
with \eqref{defY} and \eqref{eq:repr-W}  implies that
\begin{equation} \label{innyX}
    \ln X_t =
     \ln X_0 + \theta_Z(t) + \theta_B(t),
\end{equation}
where
\begin{align*}
    \theta_Z(t) &:= \rho\int_0^tY_udZ_u - \frac{1}{2}\rho^2\int_0^tY_u^2du,\\
    \theta_B(t) &:= \sqrt{1 - \rho^2}\int_0^tY_udB_u - \frac{1}{2}(1 - \rho^2)\int_0^tY_u^2du.
\end{align*}
Let $\mathcal{F}_t^{Z}= \sigma (Z_u : u\leq t)$.
For fixed $r >0$
\begin{align} \label{eq:XX}
\mathbb{P}(X_t \leq r) &= \mathbb{E}1_{\big\{X_0\exp\big(\int_0^tY_udW_u - \frac{1}{2}\int_0^tY_u^2du\big)\leq r\big\}}\\
   &= \mathbb{E}\mathbb{E}\Big[1_{\left\{\rho\int_0^tY_udZ_u +
   \sqrt{1-\rho^2}\int_0^tY_udB_u -\frac{1}{2}\int_0^tY_u^2du\leq \ln\frac{r}{X_0}\right\}}\big|\mathcal{F}_t^{Z}\Big]. \notag
 \end{align}
 Since SDE \eqref{defY} has the unique strong solution, there exists an
appropriately measurable function $\Psi(~,~)$ such that $Y=\Psi(Y_0
, Z)$. Together with the fact that the processes $B$
and $Z$ are independent Brownian motions, this implies that the random
variable $\theta_B(t)$,  for a fixed trajectory of $Z_u$, $u\leq t$,
has Gaussian distribution with mean
\[
    \hat{\mu} = - \frac{1}{2}(1 - \rho^2)\int_0^tY_u^2du
\]
and variance
\[
    \hat{\sigma}^2 = (1 - \rho^2)\int_0^tY_u^2du.
\]
Consequently, by \eqref{eq:XX}, we obtain \eqref{eq: nowy}:
 \begin{align*}
 \mathbb{P}(X_t \leq r)   &= \mathbb{E}\mathbb{P}\bigg(\mu_Z(t) + \sigma_Z(t)g\leq \ln\frac{r}{X_0}\big|\mathcal{F}_t^{Z}\bigg)\notag
    = \mathbb{E}\mathbb{P}\bigg(g\leq
    \frac{\ln\frac{r}{X_0}-\mu_Z(t)}{\sigma_Z(t)}\big|\mathcal{F}_t^{Z}\bigg) \\
    & 
    = \mathbb{E}\Phi\bigg(\frac{\ln\frac{r}{X_0}-\mu_Z(t)}{\sigma_Z(t)}\bigg), 
\end{align*}
where $\Phi$ is the cumulative distribution function of a standard
Gaussian random variable $N(0,1)$, $g$ is a standard  Gaussian random variable independent of
$\mathcal{F}_t^{Z}$,  $
    \mu_Z(t) $ and $ \sigma_Z^2(t)$ are given by \eqref{eq:repr_mu} and \eqref{eq:repr_si}, respectively.
Since
 \begin{align*}
\frac{\partial}{\partial r}
\Phi\bigg(\frac{\ln\frac{r}{X_0}-\mu_Z(t)}{\sigma_Z(t)}\bigg) =
\frac{1}{r\sigma_Z(t)}\phi\bigg(\frac{\ln\frac{r}{X_0}-\mu_Z(t)}{\sigma_Z(t)}\bigg),
\end{align*}
by Fubbini theorem for nonnegative functions, we have for $r>0$
\begin{align}
 \mathbb{P}(X_t \leq r)  & = \mathbb{E} \int_0^r \frac{1}{s\sigma_Z(t)}\phi\bigg(\frac{\ln\frac{s}{X_0}-\mu_Z(t)}{\sigma_Z(t)}\bigg) \notag \\
 & =
 \int_0^r \mathbb{E} \Big[\frac{1}{s\sigma_Z(t)}\phi\bigg(\frac{\ln\frac{s}{X_0}-\mu_Z(t)}{\sigma_Z(t)}\bigg) \Big] ds .
\end{align}
Hence the random variable $X_t$ has the
density function $g_{X_t}$ given by \eqref{eq:repr_g}. 

The continuity of density, under assumption \eqref{eq:pod}, follows from  \eqref{eq:repr_g} and the Lebesgue dominated
convergence  theorem. More precisely, we prove that 
the density $g_{X_t}$ is continuous  at  an  arbitrary  $r> 0$. 
Observe that
 $$ s\longrightarrow \frac{1}{s\sigma_Z(t)}\phi\bigg(\frac{\ln\frac{s}{X_0}-\mu_Z(t)}{\sigma_Z(t)}\bigg) $$
 is continuous on $(0, \infty)$, and
\begin{align} \label{eq:den}
\frac{1}{s\sigma_Z(t)}\phi\bigg(\frac{\ln\frac{s}{X_0}-\mu_Z(t)}{\sigma_Z(t)}\bigg) \leq \frac{1}{r-\epsilon}  \Big(\frac{1}{ \sigma_Z(t)}\Big)\phi\bigg(\frac{\ln\frac{r+\epsilon}{X_0}-\mu_Z(t)}{\sigma_Z(t)}\bigg) := J
\end{align}
 for $s \in (r-\epsilon, r+\epsilon)$. Since, by  \eqref{eq:pod},  RHS of \eqref{eq:den} (i.e. $J$) is integrable, we have   $ \lim_{s\rightarrow r}
 g_{X_t}(s) = g_{X_t}(r)$ by the Lebesgue dominated
convergence theorem.
  \end{proof}
\begin{remark} \label{uwrep}
From the last theorem it is clear that finding the
distribution of $X_t$, for fixed $t$, reduces to deriving the
distribution of the vector
$(\int_0^tY_udZ_u, \int_0^tY_u^2du)$.
\end{remark}
\begin{remark} \label{rho}
In the case of a lognormal
stochastic volatility model (i.e. in a model in which the process $Y$ is a geometric Brownian
motion) we can use the results of Matsumoto and
Yor \cite{Mat} to obtain the distribution of $(\int_0^tY_udZ_u,
\int_0^tY_u^2du)$, as we can express its components in terms of
$A_t$ and $V_t$ just as in the proof of Theorem \ref{tw:gestosc}
and use \eqref{eq:MY}.
\end{remark}
\begin{remark} \label{BS}
 Taking $Y_t \equiv
\sigma >0$ and  $\rho = 0$, we see that the Black-Scholes model
 is a linear stochastic volatility model and   Theorem
\ref{tw:densrep} gives the well known density function of a random
variable with log-normal distribution. 
\end{remark}

In the next proposition we give two sufficient conditions for \eqref{eq:pod} to hold.
\begin{proposition} \label{prop2.6}
Suppose that
 \begin{equation} \label{eq:wys-pod2}
 \ \mathbb{E}\Big(\int_0^t Y_u^2du \Big)^{-m/2} < \infty \ {\rm for}\ {\rm some}\  m\geq 1 ,
\end{equation}
or there exists $\beta>0$ and $m\geq {\frac{1}{2\beta}}$ such that
\begin{equation} \label{eq:wys-pod}
\mathbb{E} \Big( \int_0^tY_u^{-2\beta}du \Big)^m
 < \infty ,
\end{equation}
then \eqref{eq:pod} holds.
\end{proposition}
\begin{proof}
i)  Using H\"{o}lder inequality we see that \eqref{eq:wys-pod2} implies \eqref{eq:pod} for $m\geq 1$. \\
ii) Assume that \eqref{eq:wys-pod}  holds.
Since, by H\"{o}lder inequality,
\begin{equation*}
t \leq \Big( \int_0^t Y_u^{2}du \Big)^{\frac{\beta}{\beta+1}}
 \Big( \int_0^t Y_u^{-2 \beta}du \Big)^{\frac{1}{1+\beta}}   ,
\end{equation*}
we have
\begin{equation*}
 E  \Big(\int_0^t Y_u^{2}du \Big)^{-\frac{1}{2}}
\leq t^{-\frac{\beta+1}{2\beta}} E \Big( \int_0^t Y_u^{-2 \beta}du \Big)^{\frac{1}{2\beta}}   .
\end{equation*}
Hence,  using H\"{o}lder inequality with with $ p= 2 m \beta \geq 1$, we see that
\eqref{eq:wys-pod}   implies \eqref{eq:pod}.
\end{proof}

\section{Closed form of the density function in  log-normal stochastic
volatility model}

A log-normal model was considered by Hull and White  in the case
of uncorrelated noises \cite{Hul}, and it is  a SABR model with
$\beta=1$, introduced in 2002 by Hagan et al. \cite{Hag1},  in the
case of correlated noises.
 In
this case the functions  appearing in the  SDE for volatility are $\mu(y)
\equiv 0$ and $\sigma(y) = \sigma y$ for $y>0$, where $\sigma$ is a
positive constant. Thus the process $Y$ is a geometric Brownian
motion and
\begin{equation}
Y_t = Y_0 e^{\sigma Z_t - \zfrac{\sigma^2 t}{2}} \label{postacY}.
\end{equation}
Since,
\[
    \mathbb{E}\int_0^t Y_u^{-2} du = \frac{1}{3\sigma^2 Y_0^2}[e^{3\sigma^2t} - 1] < \infty,
\]
\eqref{eq:wys-pod} with $\beta=1, m=1$ is satisfied. So, by Proposition \ref{prop2.6},  a log-normal stochastic volatility model belongs to the class of
linear stochastic volatility models, which have continuous density.

Our main goal in this subsection is to find, for a log-normal
stochastic volatility model, a closed form of the density function of
the random variable $X_t$ for fixed nonnegative $t$ (see \cite{Mag1} for
another result in this direction). We determine the true
distribution of the  price process, so this allows to find a simple  way
to price derivatives in that model.

\begin{theorem} \label{tw:gestosc}
 In a log-normal
stochastic volatility model the density function of the price $X_t$  of the
underlying asset has the form
\begin{multline}
  g_{X_t}(r) = \\  \nonumber
\int_{-\infty}^{\infty}\int_{0}^{\infty}\bigg[\frac{1}{rY_0\sqrt{y\frac{1-\rho^2}{\sigma^2}}}
\phi\bigg(\frac{\ln\frac{r}{X_0}-f(x,y)+Y_0^2y\frac{1-\rho^2}{\sigma^2}}
{Y_0\sqrt{y\frac{1-\rho^2}{\sigma^2}}}\bigg)\bigg]G_{t\sigma^2}(x,y)dydx,
\end{multline}
where
\begin{align}
 \label{def-f}   f(x,y) &= \frac{\rho}{\sigma} Y_0[e^x-1]-\frac{\rho^2}{2\sigma^2}Y_0^2y,\\
\label{def-G}   G_{t}(x,y) &= \exp\bigg(-\frac{x}{2}-\frac{t}{8}-\frac{1+e^{2x}}{2y}\bigg)\theta\bigg(\frac{e^x}{y},t\bigg)\frac{1}{y},
\end{align}
and the function $\theta$ is defined, using hyperbolic functions, by
the formula
\begin{equation}
  \label{def-the}   \theta(r,t) = \frac{r}{\sqrt{2\pi^3t}}e^{\zfrac{\pi^2}{2t}}\int_0^{\infty}e^{\zfrac{-\xi^2}{2t}- r\cosh(\xi)}\sinh(\xi)\sin\bigg(\frac{\pi\xi}{t}\bigg)d\xi.
\end{equation}
\end{theorem}
\begin{proof}
Set $\tilde{Y}_t := Y_{\zfrac{t}{\sigma^2}}$.  It is clear, from
\eqref{postacY}, that
\begin{equation*}
    \tilde{Y}_t = Y_0 e^{-\zfrac{t}{2}+\tilde{Z}_t},
\end{equation*}
where $\tilde{Z}_t = \sigma Z_{\zfrac{t}{\sigma^2}}$ is a Brownian
motion. We can express $\mu_Z(t)$ and $\sigma_Z^2(t)$, defined by
\eqref{eq:repr_mu} and \eqref{eq:repr_si}, in terms of $\tilde{Y}_t
$:
\begin{equation*}
    \mu_Z(t) =  \frac{\rho}{\sigma}[\tilde{Y}_{t\sigma^2} - \tilde{Y}_0] -
    \frac{1}{2\sigma^2}\int_0^{t\sigma^2}\tilde{Y}_{u}^2du, \quad
    \sigma_Z^2(t) = \frac{1 - \rho^2}{\sigma^2}\int_0^{t\sigma^2}\tilde{Y}_{u}^2du.
\end{equation*}
Let
\begin{equation*}
 V_t := \tilde{Z}_t - \frac{t}{2}, \qquad
 A_t := \int_0^t e^{2V_s}ds.
\end{equation*}
Then $ \tilde{Y}_t = Y_0 e^{V_t}$ and $\int_0^t\tilde{Y}_{u}^2du =
Y_0^2 A_t$.
Using Theorem \ref{tw:densrep} we can write the density function
$g_{X_{\zfrac{t}{\sigma^2}}}$ in terms of $V_t$ and $A_t$:
\begin{equation} \label{eq:g1}
    g_{X_{\zfrac{t}{\sigma^2}}}(r) = \mathbb{E}\bigg[\frac{1}{rY_0\sqrt{A_t\frac{1-\rho^2}
{\sigma^2}}}\phi\bigg(\frac{\ln\frac{r}{X_0}-f(V_t, A_t)
 +Y_0^2A_t\frac{1-\rho^2}{\sigma^2}}{Y_0\sqrt{A_t\frac{1-\rho^2}{\sigma^2}}}\bigg)\bigg]  ,
\end{equation}
where $f$ is given by \eqref{def-f}. Now, we use the result of
Matsumoto and  Yor \cite{Mat} which gives the density function of the
vector $(V_t, A_t)$: they proved  that for $t>0$, $y>0$ and
$x\in\mathbb{R}$,
\begin{equation} \label{eq:MY}
\mathbb{P}(V_t\in dx, A_t \in dy) = G_t(x,y){dx dy},
\end{equation}
where
\begin{align*}
G_{t}(x,y) &= \exp\bigg(-\frac{x}{2}-\frac{t}{8}-\frac{1+e^{2x}}{2y}\bigg)\theta\bigg(\frac{e^x}{y},t\bigg)\frac{1}{y},\\
 \theta(r,t) &=
\frac{r}{\sqrt{2\pi^3t}}e^{\zfrac{\pi^2}{2t}}\int_0^{\infty}e^{\zfrac{-\xi^2}{2t}-
r\cosh(\xi)}\sinh(\xi)\sin\bigg(\frac{\pi\xi}{t}\bigg)d\xi.
\end{align*}
Hence \eqref{eq:g1} can be written in the form
 \begin{equation*} \label{eq:g2}
    g_{X_{\zfrac{t}{\sigma^2}}}(r) = \int_{-\infty}^{\infty}\int_{0}^{\infty}\bigg[\frac{1}
{rY_0\sqrt{y\frac{1-\rho^2}{\sigma^2}}}\phi\bigg(\frac{\ln\frac{r}{X_0}-f(x,y)+Y_0^2y\frac{1-\rho^2}{\sigma^2}}
{Y_0\sqrt{y\frac{1-\rho^2}{\sigma^2}}}\bigg)\bigg]G_{t}(x,y)dydx,
\end{equation*}
with $f$, $G$ given by \eqref{def-f} and  \eqref{def-G}.
Replacing $t$ by $t\sigma^2$ in the above formula  finishes the
proof.
\end{proof}

\begin{remark}
Although the  formula for the density function of the price
 in the log-normal stochastic volatility model is
complicated, this result describes the true, not approximate,
probabilistic law for $X_t$.  If $X$
is a martingale, so describes the arbitrage price of the asset,
having the density function we are able to use the risk-neutral
valuation formula to price attainable European contingent claims.
For example, evaluating the arbitrage price of  power option (see,
e.g., \cite{WY}) reduces,  by Theorem \ref{tw:gestosc}, to computing
the  integral
\[
 \int_0^{\infty}\int_{-\infty}^{\infty}\int_{0}^{\infty}\frac{[(r-K)^+]^{\alpha}}{rY_0\sqrt{y\frac{1-\rho^2}{\sigma^2}}}\Phi'\Big(\frac{\ln\frac{r}{X_0}-f(x,y)+Y_0^2y\frac{1-\rho^2}{\sigma^2}}{Y_0\sqrt{y\frac{1-\rho^2}{\sigma^2}}}\Big)G_{T\sigma^2}(x,y)dydxdr,
\]
with $f,G$ given by \eqref{def-f} and \eqref{def-G}.
 We stress that in this way we reduce the valuation problem to numerical
integration of the derived density function, as is usual in the
literature (see e.g.~\cite{Car}). Thus we avoid using asymptotic
expansions (as in \cite{Hag1}); however, some difficulties arise
during the numerical integration (see e.g.~\cite{BRY}). They are
caused by the oscillating nature of the so called Hartman-Watson
distribution density function which is a part of the density function
derived by Matsumoto and Yor \cite{Mat}.
\end{remark}

\section{Closed form  of the arbitrage price of a vanilla option in a
linear stochastic volatility model}

In this section we derive a  representation of a
vanilla option price in a linear stochastic volatility model. We are
interested in computation of the
arbitrage  prices, so the process $X$ describing the discounted
price of the asset should be a martingale. Next, as examples, we
show how to deduce from  Theorem \ref{tw:probrep} closed formulae
for option prices for the models of Section 3.
 In our examples
we give conditions guaranteeing that $X$ is a martingale.
 Then,
just as in Section 2, we show how the valuation of vanilla options
in that model can be reduced to finding the distribution of the
vector $(\int_0^tY_udZ_u, \int_0^tY_u^2du)$.

\subsection{ Representation of the arbitrage price of a vanilla option in a linear
stochastic volatility model}
Now, we provide  representations for the arbitrage
prices of European call and put options.
 These formulae
generalize the famous Black-Scholes formulae as well as
the result of Hull and White for a stochastic
volatility model with uncorrelated noises \cite{Hul}.

\begin{theorem} \label{tw:probrep}
In  a linear stochastic volatility model  the  time zero
prices of European call
and put options with strike $K>0$ and maturity $t$ have the
following  representations:
\begin{align}
            \label{eq:rep-opcja1}
& \mathbb{E} [X_t - K]^+
= X_0\mathbb{E}\big[e^{\mu_Z(t) +
\zfrac{\sigma_Z^2(t)}{2}}\Phi(d_1(t))\big]
-
K\mathbb{E}\Phi(d_2(t)), \\ & \label{eq:rep-opcja2}
\mathbb{E} [K - X_t]^+   =
K\mathbb{E}\Phi(-d_2(t))
- X_0\mathbb{E}\bigg[e^{\mu_Z(t) +
\zfrac{\sigma_Z^2(t)}{2}}\Phi(-d_1(t))\bigg],
\end{align}
 where
$$   d_1(t) = \frac{\ln\frac{X_0}{K}+\mu_Z(t)+\sigma_Z^2(t)}{\sigma_Z(t)}, \qquad
 d_2(t) = d_1(t) - \sigma_Z(t) , $$
  and $\mu_Z(t)$ and $ \sigma_Z^2(t)$ are given by \eqref{eq:repr_mu} and \eqref{eq:repr_si}.
\end{theorem}

\begin{proof}
Recall that
   $  X_t = X_0 \exp (\theta_Z(t) + \theta_B(t))$,
   where
\begin{align*}
    \theta_Z(t) &:= \rho\int_0^tY_udZ_u - \frac{1}{2}\rho^2\int_0^tY_u^2du,\\
    \theta_B(t) &:= \sqrt{1 - \rho^2}\int_0^tY_udB_u - \frac{1}{2}(1 - \rho^2)\int_0^tY_u^2du.
\end{align*}
We see that $\theta_Z(t)$ is $ \mathcal{F}^{Z}_t$-measurable, so
\begin{align*}
 \mathbb{E}(K-X_t)^+ =
  \mathbb{E}\Big[X_0 e^{\theta_Z(t)} \mathbb{E} \Big( \big(\frac{K}{X_0 e^{\theta_Z(t)}}  - e^{\theta_B(t)}\big)^+ \Big| \mathcal{F}_t^{Z}\Big)\Big] := I.
\end{align*}
We know, from the proof of Theorem \ref{tw:densrep}, that  the random
variable $\theta_B(t)$,  for a fixed trajectory of $Z_u$, $u\leq t$,
has the Gaussian distribution with mean
\[
    \hat{\mu} = - \frac{1}{2}(1 - \rho^2)\int_0^tY_u^2du = - \frac{1}{2} \sigma_Z^2(t)
\]
and variance
\[
    \hat{\sigma}^2 = (1 - \rho^2)\int_0^tY_u^2du = \sigma_Z^2(t).
\]
Using classical results we conclude that
 \begin{align*}
I & =
  \mathbb{E}\Big[X_0 e^{\theta_Z(t)}   \frac{K}{X_0 e^{\theta_Z(t)}} \Phi \Big(  \frac{- \ln\frac{X_0}{K} - {\theta_Z(t)} +\frac{\sigma_Z^2(t)}{2}}{\sigma_H}    \Big) -
 X_0 e^{\theta_Z(t)} \Phi \Big(  \frac{-\ln\frac{X_0}{K} - {\theta_Z(t)} - \frac{\sigma_Z^2(t)}{2}}{\sigma_H}    \Big)  \Big]  \\ &
=  \mathbb{E}\Big[ K  \Phi \Big(  \frac{- \ln\frac{X_0}{K} - {\mu_Z(t)} }{\sigma_H}    \Big) -
 X_0 e^{\theta_Z(t)} \Phi \Big(  \frac{-\ln\frac{X_0}{K} - {\mu_Z(t)} -  \sigma_Z^2(t) }{\sigma_H}    \Big)  \Big]  \\ & = K\mathbb{E}\Phi(-d_2(t))
- X_0\mathbb{E}\bigg[e^{{\theta_Z(t)}}\Phi(-d_1(t))\bigg].
\end{align*}
By the same arguments we have
\begin{align*}
 \mathbb{E}&(X_t-K)^+  =
  \mathbb{E}\Big[X_0 e^{\theta_Z(t)} \mathbb{E} \Big( \big(e^Z  - \frac{K}{X_0 e^{\theta_Z(t)}} \big)^+ \Big| \mathcal{F}_t^{Z}\Big)\Big] \\&
  = \mathbb{E}\Big[X_0 e^{\theta_Z(t)}
\Phi \Big(  \frac{\ln\frac{X_0}{K} + {\theta_Z(t)} + \frac{\sigma_Z^2(t)}{2}}{\sigma_H}    \Big) -  X_0 e^{\theta_Z(t)}  \frac{K}{X_0 e^{\theta_Z(t)}} \Phi \Big(  \frac{ \ln\frac{X_0}{K} + {\theta_Z(t)} - \frac{\sigma_Z^2(t)}{2}}{\sigma_H}    \Big) \Big]  \\ & =
X_0\mathbb{E}\bigg[e^{{\theta_Z(t)}}\Phi(d_1(t))\bigg] - K\mathbb{E}\Phi(d_2(t)),
\end{align*}
   which ends the proof.
\end{proof}

\begin{corollary}
 Assume that  $X$ is a martingale. Then a call-put parity holds.
\end{corollary}
\begin{proof}
Using \eqref{eq:rep-opcja1} and \eqref{eq:rep-opcja2} we have
$$
\mathbb{E}(X_t-K)^+ - \mathbb{E}(K - X_t)^+ = \mathbb{E}(X_t) - K.
$$
Hence and by the fact that $\mathbb{E}(X_t)=\mathbb{E}(X_0)$, since
$X$ is a martingale, we conclude the assertion of the corollary.
\end{proof}

\subsection{Examples}

In this subsection we consider the previously discussed models.

\subsubsection{ Black-Scholes and
 log-normal stochastic volatility models}

  In these two cases, closed formulae for the arbitrage
price of European call and put options with strike $K>0$ can be
derived.  We emphasize that these results are not a direct
consequence of deriving the density function for the model. Rather,
they are consequences of the  representation (see
Theorem \ref{tw:probrep}) of the
arbitrage price of vanilla option in a linear stochastic volatility
model.

In the case of the Black-Scholes model, $\mu_Z(t) = -\zfrac{t\sigma^2}{2}$
and $\sigma_Z^2(t) = \sigma^2t$, so \eqref{eq:rep-opcja1} and
\eqref{eq:rep-opcja2} immediately give
the famous Black-Scholes formulae.

As before, the case of a log-normal stochastic volatility model is
less trivial. We give  formulae for
the arbitrage prices of vanilla options in such models (different
formulae were obtained in \cite{Mag1} in another way).

\begin{remark} \label{uw1}
Sin \cite{Sin} and Jourdain \cite{Jou} proved that the condition $\rho\in(-1,0]$ is
equivalent to $X$ being a martingale. So, in further
considerations, whenever we need $X$ to be martingale, we consider only
nonpositive $\rho$, and in this case $\mathbb{P}$ is a martingale
measure.
\end{remark}

\begin{theorem} \label{CALLPUTFORM}
  In a log-normal stochastic volatility model
 the time zero
arbitrage prices of European call
and put
 options with strike $K>0$ and maturity $t$ are given by
\begin{multline} \label{eq:call-dokl}
    \mathbb{E}[X_t - K]^+  \\
 = \int_{-\infty}^{\infty}\int_{0}^{\infty}\bigg[X_0e^{f(x,y)}
    \Phi(d_1(x,y))-K\Phi(d_2(x,y))\bigg]G_{t\sigma^2}(x,y)dydx,
  \end{multline}
\begin{multline}  \label{eq:put-dokl}
     \mathbb{E} [K - X_t]^+  \\
 = \int_{-\infty}^{\infty}\int_{0}^{\infty}
     \bigg[K\Phi(-d_2(x,y))-X_0e^{f(x,y)}
    \Phi(-d_1(x,y))\bigg]G_{t\sigma^2}(x,y)dydx,
\end{multline}
where $f$, $G$ are given by \eqref{def-f} and \eqref{def-G},  and
\begin{align*}
    d_1(x,y) &= \frac{\ln{\frac{X_0}{K}}+f(x,y)}{Y_0\sqrt{y\frac{1-\rho^2}{\sigma^2}}}
    +\frac{Y_0}{2}\sqrt{\frac{1-\rho^2}{\sigma^2}y},\\
    d_2(x,y) &= d_1(x,y) -
    \frac{Y_0}{2}\sqrt{\frac{1-\rho^2}{\sigma^2}y}.
\end{align*}
\end{theorem}
\begin{proof}
Arguing  as in the proof of Theorem
\ref{tw:gestosc} and using the same notation we have, by  Theorem
\ref{tw:probrep},
\begin{align}
    \mathbb{E}[X_{\frac{t}{\sigma^2}} - K]^+
&=
    \mathbb{E}\big[X_0e^{f(V_t,A_t)}\Phi(d_1(V_t,A_t))-K\Phi(d_2(V_t,A_t))\big],\\
    \mathbb{E}[K - X_{\frac{t}{\sigma^2}}]^+ &=
    \mathbb{E}\big[-K\Phi(-d_2(V_t,A_t)) -
    X_0e^{f(V_t,A_t)}\Phi(-d_1(V_t,A_t))\big] ,
\end{align}
and hence
\begin{gather} \label{eq:op}
    \mathbb{E}[X_{\frac{t}{\sigma^2}} - K]^+ =
    \int_{-\infty}^{\infty}\int_{0}^{\infty}\bigg[X_0e^{f(x,y)}\Phi(d_1(x,y))-K\Phi(d_2(x,y))\bigg]G_{t}(x,y)dydx
    ,\\
 \label{eq:opp}
\mathbb{E}[K - X_{\frac{t}{\sigma^2}}]^+ =
\int_{-\infty}^{\infty}\int_{0}^{\infty}\bigg[K\Phi(-d_2(x,y))-
X_0e^{f(x,y)}\Phi(-d_1(x,y))\bigg]G_{t}(x,y)dydx.
\end{gather}
To conclude the proof we replace $t$ by $t\sigma^2$ in \eqref{eq:op}
and \eqref{eq:opp}.
\end{proof}

\section{Connection between a distribution of the asset price process and  prices  of put options}

In this section we represent the distribution of the process $X$ giving the price of the asset in a linear stochastic volatility model in terms of  prices of put options.
At first we note that $X$ is a Markov process as a strong solution to SDE \eqref{defX}.
The crucial observation in this section is that the linear stochastic volatility model has conditionally the structure of Black-Scholes model, so vanilla options prices inherit some special properties of Black-Scholes that enable us to find a probabilistic representation for a  transition density function (see Thm. \ref{tw:probrep}).
\subsection{General results}

\begin{theorem} \label{DSTR}
  In a   linear stochastic volatility model with $X_0=x$
 we have, for $r\geq 0$,
\begin{align} \label{eq:distrfunct1}
	 & \mathbb{P}(X_t \leq r) = \frac{\partial}{\partial r}\mathbb{E}_x(r - X_t)^+ , \\ &
 \label{eq:GEST-funct}
	g_{X_t}(r) = \frac{\partial^2}{\partial r^2}\mathbb{E}(r - X_t)^+dr.
\end{align}
\end{theorem}
\begin{proof} The differentiability of $r\mapsto \mathbb{E}(r - X_t)^+$ follows from  \eqref{eq:rep-opcja2} and the Lebesgue dominated
convergence theorem. Indeed, we check that the derivative of the function under expectation operator of the right side of \eqref{eq:rep-opcja2} is bounded by  integrable random variable,
so we can differentiate under expectation operator in \eqref{eq:rep-opcja2} and simple algebra leads us to
\begin{equation} \label{eq:secdistr}
	\frac{\partial}{\partial r}\mathbb{E}(r - X_t)^+ = \mathbb{E}\Phi\bigg(\frac{\ln\frac{r}{X_0}-\mu_Z(t)}{\sigma_Z(t)}\bigg),
\end{equation}
for $r>0$.
So \eqref{eq:distrfunct1}
 follows from \eqref{eq: nowy}.\\
 To prove the second part we notice that the differentiability of $r\mapsto \frac{\partial}{\partial r}\mathbb{E}(r - X_t)^+$ follows from the \eqref{eq:secdistr} and again the Lebesgue dominated
convergence theorem. This,  \eqref{eq:distrfunct1} and the existence of density imply \eqref{eq:distrfunct1}.
\end{proof}

In the next corollary we find that a Laplace transform of $X_t$ for  $\lambda > 0$ is equal to price of put option with random strike multiplies by constant.
\begin{corollary} \label{trfun} In a linear stochastic volatility model we have, for any $\lambda > 0$,
\begin{equation}
	\mathbb{E}  e^{-\lambda X_t} = \lambda \mathbb{E}(T_{\lambda} - X_t)^+,
\end{equation}
where $T_{\lambda}$ is exponential random variable with parameter $\lambda$ independent of $X$.
\end{corollary}
\begin{proof} We have, by  \eqref{eq:GEST-funct},
\begin{align} \label{eq:pom1}
	\mathbb{E} e^{-\lambda X_t} &= \int_0^{\infty}e^{-\lambda r}\frac{\partial^2}{\partial r^2}\mathbb{E}(r - X_t)^+ dr = \lambda \int_0^{\infty}\lambda e^{-\lambda r} \mathbb{E}(r - X_t)^+dr,
\end{align}
where we in the second equality we have integrated by parts and used \eqref{eq:secdistr} to conclude $\frac{\partial}{\partial r}\mathbb{E}(r - X_t)^+|_{r=0}=0$ . This is precisely the assertion of our corollary.
\end{proof}

\begin{proposition} \label{CP} If $\mathbb{E}X_t < \infty$, then for every  $r\geq 0$
\begin{equation} \label{eq:duality}
	\frac{\partial^2}{\partial r^2}\mathbb{E}(r - X_t)^+ = \frac{\partial^2}{\partial r^2}\mathbb{E}(X_t - r)^+.
\end{equation}
\end{proposition}
\begin{proof} Since
\begin{align*}
 	\mathbb{E} (X_t - r) = \mathbb{E} (X_t - r)^+ - \mathbb{E} (r-X_t)^+ ,
\end{align*}
 taking the second derivative with respect to $r$ we obtain \eqref{eq:duality}.
\end{proof}

\subsection{Log-normal stochastic volatility model}

As we mentioned in Remark \ref{uw1} 
Sin \cite{Sin} and later Jourdain \cite{Jou} proved that in the log-normal stochastic volatility model the price process $X$ is a martingale if and only if $\rho \leq 0$. Their rather technically complicated proof relied on Feller's test for explosion.
Here we have presented a simple proof of this result.
\begin{theorem} In the log-normal stochastic volatility model $X$ is a martingale if and only if $\rho \leq 0$.
\end{theorem}
\begin{proof}
Sufficiency. Take any $t\geq 0$. By \eqref{postacX}, \eqref{eq:repr-W} and \eqref{innyX} we have
\begin{align*}	
	& \mathbb{E}X_t = x\mathbb{E}e^{\int_0^tY_udW_u - \frac12\int_0^tY_u^2du} = \\& x \mathbb{E}\Big[\exp{\Big(\rho\int_0^tY_udZ_u - \frac{\rho^2}{2}\int_0^tY_u^2du\Big)} \exp{\Big(\sqrt{1 - \rho^2}\int_0^tY_udB_u - \frac{1}{2}(1 - \rho^2)\int_0^tY_u^2du \Big)
}\Big].
\end{align*}	
As processes $Y$ and $B$ are independent, we deduce taking conditional expectation and using Girsanov theorem, that \begin{align*}	
	& \mathbb{E}X_t  =  x \mathbb{E}\Big[\exp\Big(\rho\int_0^tY_udZ_u - \frac{\rho^2}{2}\int_0^tY_u^2du\Big) \Big].
\end{align*}
As the local martingale under the expectation is bounded
$$	
	e^{\rho(Y_t-Y_0) - \frac{\rho^2}{2}\int_0^tY_u^2du} \leq xe^{-\rho Y_0},
$$
it is a true martingale.
This implies that $\mathbb{E}X_t=x$ for all $t$. This concludes the proof since $X$ is a local martingale.

 Necessity.  Suppose that $\rho > 0$ and assume without loss of generality that $Y_0 = 1$. Suppose, contrary to our claim, that $X$ is a martingale. Then $M_t:= \exp\{\rho\int_0^tY_udZ_u - \frac{\rho^2}{2}\int_0^tY_u^2du\}$
 is a martingale and we  define, for $t\geq 0$, a new probability measure $Q$ by
 \begin{align*}
 \frac{d\mathbb{Q}}{d\mathbb{P}}|_{\mathcal{F}_t} := M_t.
 \end{align*}
 The process $\hat{B}_s = B_s - \rho\int_0^sY_udu$ for $s\leq t$ is a standard Brownian motion  under $\mathbb{Q}$, by the Girsanov theorem. As $Y_s = e^{B_s-s/2}$, the It\^o lemma implies
 \begin{align*}
 	0 < e^{\hat{B}_t - B_t} &= 1 + \int_0^te^{\hat{B}_u - B_u}d(\hat{B}_u - B_u)
 	= 1 - \rho\int_0^te^{\hat{B}_u - B_u}Y_udu\\ &= 1 - \rho\int_0^te^{\hat{B}_u-u/2}du.
 \end{align*}
In result,
 \begin{align*}
 		1 = \mathbb{Q}\Big(e^{\hat{B}_t - B_t} > 0  \Big) = \mathbb{Q}\Big(1 - \rho\int_0^te^{\hat{B}_u-u/2}du > 0  \Big).
 \end{align*}
Contradiction. The process $X$ can not be a martingale.
\end{proof}

In the next important example we use the notion of implied volatility in the log-normal stochastic volatility model. The implied volatility in this context is a function of three variables ( $r$ representing the exercise price, $x$ - current price of an asset and $t$ - time to expiration of an option) which inserted in the Black-Scholes price of the option gives the arbitrage price of the option in considered stochastic volatility model. But as we can see in Theorem \ref{DSTR} the second derivative of the function $r\mapsto \mathbb{E} (r - X_t)^+$ gives the  density function of distribution of the asset price $X$ in the stochastic volatility model. So putting  $\rho$ calibrated to market prices of the options we obtain the calibrated distribution of the asset price process. We formulate these consideration in the form of remark.
\begin{remark}
The log-normal stochastic volatility model is a special case of SABR model
 (parameter $\beta = 1$) for which the formula for Black--Scholes implied volatility  is given by
\begin{align*}
 	&\sigma(r,x,t) = \sigma \ln{(x/r)}\Big(1 + t(\sigma\rho y /4 + \sigma^2(2-3\rho^2)/24)\Big)\\
 	&\times\Big(\ln{\Big(\sqrt{1-2\rho\sigma\ln(x/r)/y+ (\sigma\ln(x/r)/y)^2}+ \sigma\ln(x/r)/y - \rho\Big)}-\ln(1-\rho)\Big)^{-1}
\end{align*}
(see \cite{Hag1}). In result we obtain
\begin{align*}
	\mathbb{E}(r-X_t)^+ = r\Phi(-d_2) - x\Phi(-d_1),
\end{align*}
where
\begin{align*}
d_1 =	d_1(r,x,t) = \frac{\ln( x/r) + t\sigma^2(r,x,t)/2}{\sigma (r,x,t)\sqrt{t}},\quad
d_2 =	d_2(r,x,t) = d_1(r,x,t) - \sigma (r,x,t)\sqrt{t}.
\end{align*}
This allows us to obtain, using Theorem \ref{DSTR}, the density function of $X_t$ in the Hull-White stochastic volatility model
\begin{align} \label{calibr-f}
\nonumber f(r) & =  \frac{\partial^2}{\partial r^2}\mathbb{E}(r - X_t)^+ = \frac{\partial^2}{\partial r^2}\Big(r\Phi(-d_2) - x\Phi(-d_1)\Big)\\
& = \frac{e^{-d_2^2/2}}{\sqrt{2\pi}}\Big(rd_2\Big(\frac{\partial d_2}{\partial r}\Big)^2 -2\frac{\partial d_2}{\partial r} - r\frac{\partial^2 d_2}{\partial r^2} \Big)+
\frac{xe^{-d_1^2/2}}{\sqrt{2\pi}}\Big(d_1\Big(\frac{\partial d_1}{\partial r}\Big)^2 +\frac{\partial^2 d_1}{\partial r^2} \Big) .
\end{align}
In result, when we consider the Hull-White stochastic volatility model with parameter  $\rho$ calibrated to market prices of the options, the formula \eqref{calibr-f} gives the calibrated distribution of the asset price process.
\end{remark}

\section{The Heston and extended Heston stochastic volatility models}
In this section we consider a linear stochastic volatility model with $Y^2_t = R_t$, where $R$ is a CIR or an extended CIR process. Thus, in fact, we consider the Heston stochastic volatility model and the extended Heston stochastic volatility model.
Such a model belongs to class of linear stochastic volatility models considered in this work. There is an economic motivation to model volatility of an asset by a CIR and an generalized CIR process (see for instance \cite[\& 6.3.4]{JYC}). Below we show that under martingale measure the price of an asset $X$ is always a martingale  in the  case of  a classical Heston model as well as  in the  case of an extended Heston model.
This is a new result and generalizes the results obtained by Wong and Heide \cite{HW}. We do not assume any special form of martingale measure density and do not pose any additional assumptions on model parameters. \\
Let us recall that an extended CIR process is a process $R$ given by
\begin{equation} \label{GRvol}
	dR_t = \kappa(\theta(t) - R_t)dt + \sqrt{R_t}dZ_t,
\end{equation}
where $\kappa$ is a positive constant, $\theta:[0,\infty) \mapsto [0,\infty)$ is a continuous function and $R_0 \geq 0$. It is well known that $R_t \geq 0$.
If $\theta(t)\equiv \theta >0$, then we have
 the classical CIR process given by
\begin{equation} \label{Rvol}
	dR_t = \kappa(\theta - R_t)dt + \sqrt{R_t}dZ_t,
\end{equation}
If $2\kappa\theta\geq 1$, then the process is strictly positive (see 6.3.1 in \cite{JYC}).
More properties of
CIR and extended CIR processes can be found e.g. in   \cite[Chapter 6.3]{JYC}.

\begin{remark}
If $R>0$ then we can use the It\^o lemma to write SDE for $\sqrt{R}$ and check that obtained coefficients are locally Lipschitz. Thus we obtain the linear stochastic volatility model as defined in Chapter 2, so with the volatility $Y$ given by a solution to SDE. In this case all previous results can be applied. In the general case, we can still consider linear stochastic volatility model  for $R\geq 0$ and $Y = \sqrt{R}$.
\end{remark}

 \begin{theorem} \label{HM} In the Heston and extended Heston stochastic volatility models the process $X$ is  a martingale.
\end{theorem}
\begin{proof}
For the clarity of arguments,  we divided the proof into two steps. In the first step we prove theorem for the Heston model and in the second for the extended Heston model. \\
Step 1. The  Heston model.
\\
 To prove that $X$ is a martingale it is enough to show that
\begin{align} \label{eq:1/15}
	\mathbb{E}e^{\rho\int_0^tY_udZ_u - \frac{\rho^2}{2}\int_0^tY_u^2du}=\mathbb{E}e^{\rho\int_0^t\sqrt{R_u}dZ_u - \frac{\rho^2}{2}\int_0^tR_udu}= 1.
\end{align}
For $\rho=0$ it is obvious, so we assume that $\rho\neq 0$.
 Using  \cite[Cor. 3.5.14]{KS},  a version of Novikov condition, we see that it is enough to find a monotone sequence $(t_n)$, $t_n\rightarrow\infty$, such that
\begin{align}\label{Novikov}
	\mathbb{E}e^{\frac{\rho^2}{2}\int_{t_n}^{t_{n+1}}R_udu} < \infty.
\end{align}
Define $\tilde{R}_t := R_{4t}$. Then
$$
		d\tilde{R}_t = 4\kappa(\theta - \tilde{R}_t)dt + 2\sqrt{\tilde{R}_t}d\tilde{Z}_t,
$$
where $\tilde{Z}$ is a standard Brownian motion. From comparison theorem for SDE's ~\cite[Prop. 5.2.18]{KS}) $\tilde{R}_t\leq G_t$, where $G_0 = \tilde{R}_0$ and
$$
	dG_t = 4\kappa\theta dt + 2\sqrt{G_t}d\tilde{Z}_t,
$$
so $G$ is a squared Bessel process. This means there exists an $M\in\mathbb{N}$ such that
\begin{equation} \label{estm}
G_t \leq (B_1(t) + G_0)^2 + \sum_{i=2}^M B^2_i(t)\leq 2G_0^2 + 2B_1^2(t) +  \sum_{i=2}^M B^2_i(t),
\end{equation}
where $B_i$ are the independent standard Brownian motions.
Hence, by independence of random variables on the RHS of \eqref{estm}, it is enough to prove \eqref{Novikov} for $2B_1^2$ instead of $R$.
Let us observe that  for an arbitrary $t\geq 0$ and $s \in (0, \frac{1}{2}\sqrt{t^2+2/\rho^2}-t)$
\begin{align} \label{eq:t*}
	\mathbb{E}e^{2\frac{\rho^2}{2}\int_t^{s+t}B^2_1(u)du} < \infty.
\end{align}
Indeed, for a fixed $t\geq 0$ and $s$  such  that $0< s < \frac{1}{2}(\sqrt{t^2+2/\rho^2}-t)$,
we obtain
\begin{align*}
	\int_t^{t+s}\mathbb{E}e^{\rho^2sB_1^2(u)}du < \infty,
\end{align*}
by properties of gaussian  distribution. By Jensen inequality we have
\begin{align*}
	\mathbb{E}e^{\rho^2\int_t^{t+s}B^2_1(u)du} \leq \frac{1}{s}\int_t^{t+s}\mathbb{E}e^{\rho^2t^*B_1^2(u)}du,
\end{align*}
so \eqref{eq:t*} holds. Now, we  define a sequence $t_n\rightarrow\infty$ such that \eqref{Novikov} for $B_1^2$ instead of $R$ holds. Observe that for $t > \sqrt{\frac{\rho^2}{2(1-\rho^2)}}$ we have
\begin{align}\label{j2}
\frac{1}{2t} < \frac{1}{2}(\sqrt{t^2+2/\rho^2}-t).
\end{align}
 Let  $\hat{t} =\sqrt{\frac{\rho^2}{(1-\rho^2)}}$
 and
$t^* = \frac{\sqrt{1-\rho^2}}{|2\rho^3+\rho|}$. At first, assume that $\hat{t} > t^*$. For any $u \leq \hat{t} - t^*$ we have
\begin{align}\label{j1}
t^*(t^*+u)\leq t^*\hat{t} = \frac{\sqrt{1-\rho^2}}{|2\rho^3+\rho|} \sqrt{\frac{\rho^2}{(1-\rho^2)}} = \frac{1}{2 \rho^2 + 1}<\frac{1}{2\rho^2},
\end{align}
 which in turn implies that
$t^* < \frac{1}{2}(\sqrt{u^2+2/\rho^2}-u)$. Using these observation we define  a sequence  $(t_n)_n$.
Let
 $n_0 = \inf\{k\in\mathbb{N} : (k+1)t^*\geq \hat{t} \, \}$.
 Put $t_0 = 0, t_1 = t^*, t_2 = 2t^*,...,  ,
 t_{n_0} = n_0 t^*, t_{n_0+1}=\hat{t}$ and $t_{k+1} = t_k+\frac{1}{2t_k}$ for $k \geq n_0+1$.
 We have  $0< t_{n+1} - t_n < \frac{1}{2}(\sqrt{t_n^2+2/\rho^2}-t_n)$, by definition of $(t_n)$ and \eqref{j1} for $n\leq n_0$, and \eqref{j2} for $n > n_0$. Thus   (\ref{eq:t*}) is satisfied for each $n$. Moreover, $t_n \rightarrow\infty$.
Indeed, $t_n$ is monotone, so $\lim_{n\rightarrow\infty} t_n = g $ exists. If $g<\infty$, then $g = g + \frac{1}{2g}$, by the definition of $t_n$. Contradiction.
Next, if
$\hat{t} \leq t^*$, then $2\rho^4 \leq 1 - 2\rho^2 < 1 -\rho^2$
which implies that $\hat{t}< \frac12\sqrt{\frac{2}{\rho^2}}$. Therefore \eqref{eq:t*} is satisfied for  $t = 0$ and $s = \hat{t}$. Thus,  as a desired sequence we can take $t_0 = 0$, $t_1 = \hat{t}$ and $t_{k+1} = t_k+\frac{1}{2t_k}$ for $k \geq 1$.
In result (\ref{Novikov}) is satisfied, and the proof of the first step is complete from \cite[Cor. 3.5.14]{KS}.
\\
Step 2. The extended Heston model. We follow the idea of Step 1.
\\
Again,
it is enough to show that for an extended CIR  process $R$ equality \eqref{eq:1/15} holds.
Define $\tilde{R}_t := R_{4t}$. Then
$$
		d\tilde{R}_t = 4\kappa(\tilde\theta(t) - \tilde{R}_t)dt + 2\sqrt{\tilde{R}_t}d\tilde{Z}_t,
$$
where $\tilde\theta(t) = \theta(4t)$ and $\tilde{Z}$ is a standard Brownian motion. From comparision theorem for SDE's ~\cite[Prop. 5.2.18]{KS} $\tilde{R}_t\leq G_t$, where $G_0 = \tilde{R}_0$ and
$$
	dG_t = 4\kappa\tilde\theta(t) dt + 2\sqrt{G_t}d\tilde{Z}_t.
$$
Since $\theta$ is continuous, for every $n$
 there exists a constant  $M = M(n)\in\mathbb{N}$ such that
 $\tilde\theta(\cdot)\leq M$ on $[n, n+1]$ and
\begin{equation*} \label{gestm}
G_t \leq (B_1(t) + G_0)^2 + \sum_{i=2}^M B^2_i(t)\leq 2G_0^2 + 2B_1^2(t) +  \sum_{i=2}^M B^2_i(t).
\end{equation*}
For every $n$, using the first step, we have a finite set ${\mathcal T}_n$ of points $t^{(n)}_i$ such that
$n=t_1^{(n)} < \ldots < t^{(n)}_{n_k}=n+1$ and \eqref{Novikov}
 holds.
Arranging all elements of $\bigcup_{n=1}^\infty {\mathcal T}_n$ in the incresing sequence
 finishes the proof.
\end{proof}
From Theorem \ref{HM} we know that the first moment of $X_t$ exists.
Our next goal is to give conditions ensure that the $k$-moment of the $X$ in the Heston stochastic volatility model exists.
\begin{proposition}\label{HestonEX}
Let $\rho\leq 0$. If the natural number  $k$ satisfies $k \leq \frac{1}{1-\rho^2}$, then the $k$-moment of $X_t$ exists for $t\geq 0$ in the Heston and extended Heston models.
\end{proposition}
\begin{proof} Fix  $t\geq 0$. It is enough to  prove the existence of moment for the extended Heston model. From \eqref{postacX}, from the fact $R = Y^2$ and from (\ref{Rvol}) we have
\begin{align*}
	\mathbb{E}X_t^k &= x^k\mathbb{E}e^{k\int_0^tY_udW_u - \frac{k}{2}\int_0^tY_u^2du} = x^k\mathbb{E}e^{k\rho\int_0^tY_udZ_u - \Big(\frac{k}{2}-\frac{k^2(1-\rho^2)}{2}\Big)\int_0^tY_u^2du}.
\end{align*}
 By \eqref{GRvol}
 $$\int_0^tY_udZ_u = Y_t^2 - Y_0^2 - \kappa\int_0^t\theta(u)du +\kappa\int_0^tY_u^2du = R_t - R_0 - \kappa\int_0^t\theta(u)du  +\kappa\int_0^tR_udu$$
  and $R_t\geq 0$. In result
\begin{align} \label{postac-mom-X}
	\mathbb{E}X_t^k &= x^k e^{-k\rho R_0-k\rho\kappa\int_0^t\theta(u)du}\mathbb{E}e^{k\rho R_t +k\rho\kappa\int_0^tR_udu- \Big(\frac{k}{2}-\frac{k^2(1-\rho^2)}{2}\Big)\int_0^tR_udu}\\
\nonumber		&\leq x^k e^{-k\rho R_0-k\rho\kappa\int_0^t\theta(u)du},
\end{align}
because $R_s \geq 0$, $\rho\leq 0$ and $k(1-\rho^2) \leq 1.$
The result follows.
\end{proof}

\begin{remark}
Formula \eqref{postac-mom-X} gives a form of the  $k$-moment of $X$ in terms of  the Laplace transform $\mathbb{E}e^{-\lambda R_t -\gamma \int_0^tR_udu}$ for $\lambda\geq0$ and $\gamma > 0$. For the  CIR process  the form of this transform is well known (see e.g. Proposition 6.3.4.1 in \cite{JYC}).
In the next theorem we generalize this result and  present an explicite form of Laplace transform
for
an extended CIR process.
This,  in particular, enables us to use \eqref{postac-mom-X} to find an explicite form of the  $k$-moment of $X$.
\end{remark}

\begin{theorem} \label{GHestonL} Let $R$ be an extended CIR process.  For $\lambda\geq 0,\ \gamma > 0,\ t\geq 0$ $\lambda > \sqrt{\kappa^2+2\gamma}-\kappa$  
  we have
\begin{align} \label{eq-tL-R}
\mathbb{E}e^{-\lambda R_t -\gamma \int_0^tR_udu} = e^{-R_0f(t)-\kappa\int_0^t\theta(s)f(s)ds},
\end{align}
where
\begin{align} \label{def:f1}
f(t) &= \frac{\kappa+\sqrt{\kappa^2+2\gamma}+ce^{\sqrt{\kappa^2+2\gamma}t}(\sqrt{\kappa^2+2\gamma}-\kappa)}{ce^{\sqrt{\kappa^2+2\gamma}t}-1},\\
\label{def:c1}
c &= \frac{\lambda +\kappa +\sqrt{\kappa^2+2\gamma}}{\lambda +\kappa -\sqrt{\kappa^2+2\gamma}}>1.
\end{align}
\end{theorem}
\begin{proof} Let us denote $R_0 = r >0$. Define $p_{\gamma}(t,\lambda) := \mathbb{E}e^{-\lambda R_t -\gamma \int_0^tR_udu}$ for $\lambda\geq 0,$ $t\geq 0$. Using the It\^o lemma we obtain
\begin{align}\label{Geito}
	d e^{-\lambda R_t -\gamma \int_0^tR_udu} &= -\lambda e^{-\lambda R_t -\gamma \int_0^tR_udu}\Big(\sqrt{R_t}dZ_t +\kappa(\theta(t)-R_t)dt\Big)\\
	&-\gamma e^{-\lambda R_t -\gamma \int_0^tR_udu} R_tdt +\frac12e^{-\lambda R_t -\gamma \int_0^tR_udu}\lambda^2R_tdt.\notag
\end{align}
As $e^{-\lambda R_t -\gamma \int_0^tR_udu}\leq 1$ and as for fixed $t>0$ the function $\sup_{u\leq t}\theta(u) < M$ for some $M\in\mathbb{N}$, we can use the same idea as Theorem \ref{HM} (see formula (\ref{estm})) and conclude the local martingale on the right side of (\ref{Geito}) is a martingale.
 Thus  taking expectation
 in
(\ref{Geito}) we obtain
\begin{align} \label{Geito2}
	\frac{\partial p}{\partial t} &= \Big(\gamma - \kappa\lambda -\frac{\lambda^2}{2}\Big)\frac{\partial p}{\partial\lambda} - \lambda\kappa\theta(t) p,\\
	p&(0,\lambda) = e^{-\lambda r}.\notag
\end{align}
Let us consider a diffusion $U$ (in fact a deterministic one) given by
\begin{equation}\label{Geuu}
	dU_t = (\gamma - \kappa U_t - \frac12 U_t^2)dt
\end{equation}
with $U_0 = \lambda$. The coefficient in \eqref{Geuu} is locally Lipschitz, so there exists the unique solution.
In what follows we give an explicite form of nonexploding solution to (\ref{Geuu}).
Observe that $U_t \geq 0$, again by comparision criterion for SDE (see ~\cite[Ex. 2.19, Chapter V]{KS}, if $b_1(x) = - \kappa x - \frac12 x^2$ then the unique solution of
$dU_t = b_1(U_t)dt,$ $U_0=0$
is a function identically equal to $0$) and $b_1(x) < \gamma- \kappa x - \frac12 x^2$). Let us define
$q(t,\lambda):= \kappa\theta(t)\lambda$ and consider the Cauchy problem
\begin{align} \label{GCP}
	&\frac{\partial \tilde{p}}{\partial t} = \mathcal{A}_U\tilde{p} - q\tilde{p}\\
	& \tilde{p}(0,\lambda) = e^{-\lambda r} ,\notag
\end{align}
where $\mathcal{A}_U$ is the generator of $U$.
  The function $p$  is a solution of (\ref{GCP}), since $p$  satisfies \eqref{Geito2}. From the Feynman-Kac theorem and from the fact that $U$ is deterministic we obtain that
\begin{equation}\label{form}
	p(t,\lambda) = e^{-rU_t - \kappa\int_0^t\theta(s)U_sds} .
\end{equation}
So to conclude the proof we have to find the explicit form of $U$. Therefore, we have to  solve the ordinary differential equation given by (\ref{Geuu}). Assume for the moment that $U_t+\kappa\neq \sqrt{\kappa^2+2\gamma}$ for all $t$. We have
\begin{align*}
	\frac{dU_t}{\gamma - \frac{U_t^2}{2}-\kappa U_t} = dt
\end{align*}
and from that
\begin{align}\label{ut}
	t\sqrt{\kappa^2+2\gamma} + c^* = \ln \frac{U_t+\kappa+\sqrt{\kappa^2+2\gamma}}{|U_t+\kappa-\sqrt{\kappa^2+2\gamma}|}.
\end{align}
Since $U_0 = \lambda$ we obtain
\begin{align*}
	c := e^{c^*} = \ln \frac{\lambda+\kappa+\sqrt{\kappa^2+2\gamma}}{\lambda+\kappa-\sqrt{\kappa^2+2\gamma}} > 1.
\end{align*}
Let us assume that $U_t+\kappa> \sqrt{\kappa^2+2\gamma}$.
Then, by (\ref{ut}),
\begin{align*}
	ce^{\sqrt{\kappa^2+2\gamma}t} = \frac{U_t+\kappa+\sqrt{\kappa^2+2\gamma}}{U_t+\kappa-\sqrt{\kappa^2+2\gamma}}
\end{align*}
and
\begin{align}\label{finalu}
	U_t = \frac{\kappa+\sqrt{\kappa^2+2\gamma}+ce^{\sqrt{\kappa^2+2\gamma}t}(\sqrt{\kappa^2+2\gamma}-\kappa)}
{ce^{\sqrt{\kappa^2+2\gamma}t}-1}.
\end{align}
Thus, $U$ given by \eqref{finalu} is the unique solution to the differential equation  (\ref{Geuu}) and satisfies $U_t >\sqrt{\kappa^2+2\gamma}-\kappa.$ This concludes the proof.
\end{proof}
Using  Theorem \ref{GHestonL} we obtain an alternative
proof of the well-known result for a classical CIR process (\cite[Prop. 6.3.4.1]{JYC}).
 \begin{corollary}
 For a classical CIR process $R$ and for $\lambda\geq 0,\ \gamma \geq 0,\ t\geq 0$ $\lambda > \sqrt{\kappa^2+2\gamma}-\kappa$  we have
\begin{align*}
\mathbb{E}e^{-\lambda R_t -\gamma \int_0^tR_udu} = e^{-R_0f(t)+\theta\kappa t(\kappa+\sqrt{\kappa^2+2\gamma})}\Big(ce^{\sqrt{\kappa^2+2\gamma}t}-1\Big)^{-2\kappa\theta},
\end{align*}
where $f$ is  given by \eqref{def:f1} and $c$ is  given by  \eqref{def:c1}.
\end{corollary}
\begin{proof}
In  a classical CIR process  $\theta(t)\equiv \theta >0$. Therefore to prove corollary it is enough to
find $\int_0^tU_sds$ for $\theta(t)\equiv \theta >0$ and $U$ given by  \eqref{finalu}. Observe that for constants $A >0,\ B>0,\ C>1, \ D>1$
\begin{align*}
	\int\frac{A+Be^{Cu}}{De^{Cu}-1}du &= \int\frac{A+Bv}{Dv-1}\frac{1}{Cv}dv = \frac{AD+B}{CD}\ln(Dv-1)-\frac{A}{C}\ln v,
\end{align*}
where $v=e^{Cu}$. Thus we have
\begin{align*}
	\int_0^tU_sds &= 2\ln \Big(ce^{t\sqrt{\kappa^2 +2\gamma}}-1\Big) - (\kappa + \sqrt{\kappa^2+2\gamma})t.
\end{align*}
After inserting the last result in (\ref{eq-tL-R}) we finish the proof.
\end{proof}

\begin{remark}
From Theorem \ref{GHestonL} we can obtain the density of the vector \\
$(\int_0^tY_udZ_u,\int_0^tY_u^2du)$.
Indeed, Theorem \ref{GHestonL} gives us, for a fixed $t\geq0$, the Laplace transform of $(R_t,\int_0^tR_udu) = (Y_t^2,\int_0^tY_u^2du)$.
Inverting (for instance numerically) the  Laplace transform we obtain the density of vector $(Y_t^2,\int_0^tY_u^2du)$.
 For the extended Heston stochastic volatility model we have
$$\int_0^tY_udZ_u = Y_t^2 - Y_0^2 - \kappa\int_0^t\theta(u)du +\kappa\int_0^tY_u^2du.$$
All these facts together give us numerically the form of density  of $X_t$ (see Theorem \ref{tw:densrep}).  
\end{remark}

\begin{remark}
 We can approximate the  price of put option in an extended Heston model in the case  $\rho \leq 0$ using Corollary \ref{trfun} and  Theorem \ref{GHestonL}. Indeed, for $\lambda > 0$ we have
\begin{align}\label{LamX}
	\int_0^{\infty}e^{-\lambda u}\mathbb{E}(u - X_t)^+du = \frac{1}{\lambda^2}\mathbb{E}e^{-\lambda X_t},
\end{align}
by \eqref{eq:pom1}. If $\rho \leq 0$ and $n \leq \frac{1}{1-\rho^2}$ for $i \leq n$ we can compute $\mathbb{E}X_t^i$ using Theorem \ref{GHestonL} (see \eqref{postac-mom-X}).
Now, we use the following approximation $$\mathbb{E} e^{-\lambda X_t} \approx \sum_{i=0}^{n}\frac{(-\lambda)^i}{i!}\mathbb{E}X_t^i.$$
In result from (\ref{LamX}) we have
\begin{align}\label{LamX-2}
\int_0^{\infty}e^{-\lambda u}\mathbb{E}(u - X_t)^+du\approx \frac{1}{\lambda^2}\sum_{i=0}^{n}\frac{(-\lambda)^i}{i!}\mathbb{E}X_t^i = \sum_{i=0}^{n}\frac{(-\lambda)^{i-2}}{i!}\mathbb{E}X_t^i .
\end{align}
Now to find the approximate price of the put option $\mathbb{E}(u - X_t)^+$ we have to find the invert Laplace transform (at least numerically) of the left hand side of \eqref{LamX-2}.
\end{remark}

\bibliographystyle{plain}

\end{document}